\documentclass{article}
\usepackage{slashed,amsmath,amssymb,enumitem}
\usepackage[papersize={8.5in,11in}]{geometry}
\begin{document}
\title{Quarks and Bosons as Composite Particles}
\author{J. W. Moffat\\~\\
Perimeter Institute for Theoretical Physics, Waterloo, Ontario N2L 2Y5, Canada\\
and\\
Department of Physics and Astronomy, University of Waterloo,\\
Waterloo, Ontario N2L 3G1, Canada}
\date{\today}
\maketitle
\begin{abstract}%
A composite model of quarks and bosons is proposed in which a spin $1/2$ isospin doublet $\psi$ is the basic building block of quarks, $W^\pm$, $Z^0$ and Higgs boson $H^0$ in the standard model. The $\psi$ has two components $\alpha$ and $\beta$ with charges $Q_\alpha=\frac{2}{3}e$ and $Q_\beta=-\frac{1}{3}e$, respectively. The two constituent flavors $\alpha$ and $\beta$ combine to form the three generations of colored standard model quark flavors. A triplet of massless hypergluons binds the constituents called geminis. The strong confining non-Abelian $SU(2)_C$ constituent color dynamics has a confining energy scale $\Lambda_{CD}$. The leptons, the massless photon, gluons and triplet of hypergluons are treated as point particles. The non-Abelian $SU(2)_C$ color dynamics satisfies asymptotic freedom, and, together with the compositeness of the Higgs boson $H^0$, resolve the Higgs mass and gauge hierarchy fine-tuning problems. 
\end{abstract}


\section{Introduction}

In a previous publication, a composite model of quarks and bosons was introduced~\cite{Moffat}\footnote[1]{A composite model of the weak bosons $W^\pm$ and $Z^0$ has been proposed by Harold Fritzsch~\cite{Fritzsch}.}.  We will replace this model with a new one that also explains the compositeness of standard model (SM) particles in terms of basic $SU(2)$ flavor and colored doublets with a non-Abelian confining color group $SU(2)_C$. The $SU(2)$ constituents of the quarks and bosons are called geminis\begin{footnote}[2]{Gemini is the Latin word for twin. They are not identical twins.}\end{footnote} and they are strongly confined. For the energy scale $E\lesssim E_c$, the SM quarks and $W^\pm$, $Z^0$ and Higgs boson $H^0$ will be observed at the LHC as elementary point particles due to the strong confinement of the constituent gemini particles. The  leptons and the massless photon, $SU(3)$ colored gluons and $SU(2)_C$ hypergluons are treated as elementary point particles. 

The problems of the fine-tuning of the Higgs boson mass and the gauge hierarchy are produced by the quadratic divergence in the radiative correction to the scalar boson bare mass and that $v_H\ll\Lambda$ where $v_H=246$ GeV and $\Lambda$ is an energy cutoff~\cite{Wilson,tHooft}. The Higgs boson self-energy $\delta m_H^2\sim\Lambda^2/16\pi^2$ leads to large corrections to the Higgs mass for an arbitrarily large $\Lambda$.  Moreover, the electroweak (EW) vacuum can be critically metastable at a high energy scale~\cite{Branchina}. Because the SM has more than 20 free parameters and the potential exists for new physics to be discovered at the LHC and at future accelerators, the SM can only be considered an incomplete theory subject to hierarchy problems and fine-tuning. The composite model can resolve the Higgs mass and gauge hiererarchy problems of the SM model~\cite{Moffat}. 

The LHC has not detected any composite substructure of the standard model fermions and bosons. Arguments make us believe that possible substructures will only be observed at very short distances and large momenta, which are expected to occur for length scales less than $10^{-16}\,{\rm cm}$. Perhaps, at a future supercollider, we can reach a length scale at which we can detect the composite nature of particles. 

\section{Composite Model}

The most economical model for the basic building blocks of matter is based on the spin 1/2 iso-doublet $\psi=\begin{pmatrix}\alpha\\\beta\end{pmatrix}$. The constituent component gemini particles $\alpha$ and $\beta$ charges are $Q_\alpha=\frac{2}{3}e$ and $Q_\beta=-\frac{1}{3}e$, respectively. The up and down quarks are formed from $u(\alpha\beta\bar\beta)$ and $d(\alpha\beta\bar\alpha)$, respectively. The isospin quantum number $I_3=1/2$ for $u(\alpha\beta\bar\beta)$ and $I_3=-1/2$ for $d(\alpha\beta\bar\alpha)$. The other flavors of quarks have $I_3=0$ and  are formed from $s(\alpha\beta\bar\alpha)$, $c(\alpha\beta\bar\beta)$, $b(\alpha\beta\bar\alpha)$, $t(\alpha\beta\bar\beta)$ and they have the quantum numbers $S=-1$ for the strange quark $s$, C=+1, for the charm quark $c$, $B'=-1$ for the bottom quark $b$, and $T=+1$ for the top quark $t$. The quark charge $Q_q$ is given by
\begin{equation}
Q_q=I_3+\frac{1}{2}(B+S+C+B'+T),
\end{equation}
where $B=1/3$ denotes the quark baryon number. 

We treat the leptons, $e,\mu,\tau,\nu_e,\nu_\mu,\nu_\tau$ as point particles\begin{footnote}[3]{We leave open the option of also treating the leptons as composites of the gemini particles.}\end{footnote}. As in the case of the octet of gluons in $SU(3)_C$ that confines the SM quarks the $SU(2)_C$ triplet of hypergluons are electrically neutral and massless. The spin 1 gauge bosons $W^{\pm}$ and $Z^0$ are formed from combinations of the $\alpha$ and $\beta$ quarks: $W^+(\alpha\bar\beta), W^-(\bar\alpha\beta)$ and $Z^0(\alpha\bar\alpha-\beta\bar\beta)$, while the composite scalar Higgs boson $H^0$ is formed from 
$ H^0(\alpha\bar\alpha+\beta\bar\beta)$. The neutral left and right components of the gauge boson $W^3_\mu$ are given by 
\begin{equation}
W^3_{\mu L,R}=\frac{1}{\sqrt{2}}({\alpha}_{L,R}\gamma_\mu\bar\alpha_{L,R}+{\beta}_{L,R}\gamma_\mu \bar\beta_{L,R}). 
\end{equation}
The neutral singlet gauge boson $B_\mu$ is orthogonal to $W^3_{\mu L}$ and $W^3_{\mu R}$: 
\begin{equation}
B_\mu=\frac{1}{2}({\alpha}_{L}\gamma_\mu \bar\alpha_{L}+{\alpha}_{R}\gamma_\mu \bar\alpha_{R}-{\beta}_{L}\gamma_\mu\bar\beta_{L}-{\beta}_{R}\gamma_\mu\bar\beta_{R}).
\end{equation}

A dynamical model of the constituent $\alpha$ and $\beta$ and the hypergluons can be constructed that binds the constituents in the SM quarks, while $SU(3)_C$ binds the quarks in the hadrons. The $\alpha$ and $\beta$ have two colors belonging to the representation $\underline 2$ of $SU(2)_C$ and we have $\underline 2\times \underline 2^*=\underline 1+\underline 3$. The triplet $\underline 3$ of hypergluons couple to the two colors red and yellow, $\alpha_R,\alpha_Y$ and $\beta_R,\beta_Y$. The non-Abelian $SU(2)_C$ hypergluons and the strong constituent dynamics (CD) confine the colored geminis inside the quarks, which in turn carry the three $SU(3)_C$ red, blue and green color charges for the $u$ and $d$ quarks and the additional flavor generations of colored SM $s,c,b$ and $t$ quarks. As in the SM the hadrons are formed from the singlet (white) color combinations of quarks. 

We adopt the $SU(2)_C$ non-Abelian color charge Lagrangian:
\begin{equation}
\label{CDLagrangian}
{\cal L}_{CD}={\bar \chi}_i(i\slashed\partial-m_{\chi_i})\chi_i-ig_{CD}{\bar\chi_i}(\gamma^\mu T_a{\tilde G}^a_\mu)\chi_i-\frac{1}{4}{\tilde G}^a_{\mu\nu}{\tilde G}^{\mu\nu}_a,
\end{equation}
where the Dirac spinors $\chi_i$ have internal flavor and color charge components, $\slashed\partial$ denotes $\gamma^\mu\partial_\mu$, $T_a$ are the generators of the group $SU(2)_C$, $g_{\rm CD}$ is the constituent dynamic coupling constant and ${\tilde G}^a_\mu$ denotes the hypergluon triplet of fields.  
For vanishing masses $m_{\chi_ i}=0$, the Lagrangian (\ref{CDLagrangian}) with $\chi_{iL},\chi_{iR}$ for the constituent Dirac fields possesses a scale invariant and chiral symmetry. The scale invariance is broken by the constituent confinement energy scale $\Lambda_{CD}$. We can anticipate that the CD confining $SU(2)_C$ coupling constant $\alpha_{CD}=g^2_{CD}/4\pi$ will be screened and lead to asymptotic freedom. This will cause a damping of high energy momentum processes and lead to a UV complete model, analogous to the flavor color screening and asymptotic freedom of QCD. 

For the gauge color group $SU(N)_C$ the beta function in lowest order perturbation theory is given by
\begin{equation}
\beta(g_{CD})=-\frac{g_{CD}^3}{48\pi^2}\biggl(11C_2-\frac{4}{3}N_fC(R)\biggr),
\end{equation}
where $C_2$ is the quadratic Casimir coefficient of $SU(N)_C$ and $C(R)$ is the Casimir invariant defined by $Tr(T^a_RT^b_R)=C(R)\delta^{ab}$ for the group generators $T_R^{a,b}$ of the Lie alegbra in the representation $R$. For the gluons in the adjoint representation of $SU(N)_C$, we have $C_2=N_C$ (where $N_C$ is the number of colors) and for fermions in the fundamental representation: $C(R)=1/2$.  For $SU(3)_C$ in QCD, $N_C=C_2=3$ and for the constituent dynamics $SU(2)_C$, $N_C=C_2=2$. 
The running CD constituent coupling constant $\alpha_{CD}(q^2)$ for the color gauge group $SU(2)_C$ is given by
\begin{equation}
\label{runningalpha}
\alpha_{CD}(q^2)=\frac{\alpha_{CD}(\mu^2)}{1+\frac{\alpha_{CD}(\mu^2)}{12\pi}(22-2N_{cf})\ln(q^2/\mu^2)},
\end{equation}
where $\mu$ is a renormalization group flow mass and the number of constituent flavors $N_{cf}=2$.  From (\ref{runningalpha}) we see that at sufficiently low $q^2$, the effective coupling constant will become large.  At the $q^2$ scale at which this happens the constituent confinement scale $\Lambda_{CD}$ is given by
\begin{equation}
\Lambda_{CD}^2=\mu^2\exp\biggl[-\frac{12\pi}{(22-2N_{cf})\alpha_{CD}(\mu^2)}\biggr].
\end{equation}
It follows that (\ref{runningalpha}) can be written as 
\begin{equation}
\label{alpha}
\alpha_{CD}(q^2)=\frac{12\pi}{(22-2N_{cf})\ln(q^2/\Lambda_{CD}^2)}.
\end{equation}
For $q^2\gg \Lambda_{CD}^2$, the effective coupling is small and this allows for a perturbative description of the geminis and hypergluons interacting weakly, while for $q^2\sim\Lambda_{CD}^2$, we cannot use perturbation theory and the geminis arrange themselves into strongly bound SM quarks. We can think of $\Lambda_{CD}$ as marking the boundary between quasi-free geminis and hypergluons, and quarks, gluons and hadrons. The value of $\Lambda_{CD}$ must be determined by experiment.

A natural solution to the gauge hierarchy problem can be obtained, as in standard QCD, through asymptotic freedom~\cite{Gross,Politzer} associated with the exact non-Abelian $SU(2)_C$ symmetry. A UV fixed point at which $\alpha_{CD}$ tends to zero according to (\ref{alpha}):
\begin{equation}
\alpha_{CD}(q^2)\propto \frac{1}{\ln(q^2/\Lambda^2_{\rm CD})},
\end{equation}
yields a natural cutoff for quantum corrections, which are kept under control. Because $\alpha_{CD}(\Lambda_{\rm PL})\ll1$ at the Planck energy scale $\Lambda_{\rm PL}$ an exponential suppression leads to
\begin{equation}
\Lambda^2_{\rm CD}\sim\exp\biggl(-\frac{1}{\alpha_{CD}(\Lambda_{\rm PL}^2)}\biggr)\Lambda^2_{\rm PL}.
\end{equation}
This solves the gauge hierarchy problem, and our composite model of quarks and bosons can describe a theory which is free of the SM fine-tuning problems and is UV complete.

\section{Electroweak Symmetry Breaking}

We have two possible choices to explain electroweak symmetry breaking. The first is to adopt the standard Higgs spontaneous symmetry breaking with 
$v_H=\langle\phi\rangle_0\neq 0$ where $\phi$ denotes the 125 GeV Higgs field. This leads in the familiar way to the prediction of the $W^\pm$ and $Z^0$ masses. A second choice is to assume that the constituent doublet $\psi=\begin{pmatrix}\alpha\\\beta\end{pmatrix}$ engages in electroweak as well as strong interactions~\cite{Moffat}. A chiral condensate: 
\begin{equation}
\label{condensate}
v_{CD}=\langle\bar\alpha_L\alpha_R+\bar\beta_L\beta_R\rangle_0\neq 0
\end{equation}
will be generated, which spontaneously breaks the chiral symmetry $SU(2)_R\times SU(2)_L$ of the Lagrangian\begin{footnote}[4]{A similar spontaneous symmetry breaking mechanism is introduced in QCD and technicolor~\cite{Susskind}}\end{footnote}. We turn on the $SU(2)_L\times U(1)_Y$ electroweak interaction without the scalar fields usually introduced to give masses to the $W_\mu^a$ associated with $SU(2)_L$  and the $B_\mu$ associated with weak hypercharge $Y$. Three massless Nambu-Goldstone bosons $\sigma^a$ play the role of the massless scalar fields and become the longitudinal components of the massive $W_\mu^{\pm}$ and $Z_\mu^0$.  

The $CD$ condensate (\ref{condensate}) spontaneously breaks the EW gauge dynamics, and the Goldstone theorem says that there will be a massless spin 0 boson for each broken generator with three massless Nambu-Goldstone bosons forming an isotriplet. The constituent particle decay constant $f_{CD}$ is defined by
\begin{equation}
\langle 0\vert J^{\pm}_\mu\vert \sigma^{\pm}\rangle=f^{\pm}_{CD}q_\mu,
\end{equation}
and
\begin{equation}
\langle 0\vert J^Y_\mu\vert \sigma^0\rangle=f^0_{CD}q_\mu,
\end{equation}
where $J^{\pm}_\mu$ are two $SU(2)_L$ currents and $J^Y_\mu$ is the hypercharge current. Because of the conservation of electromagnetic charge, $f^+_{CD}=f^-_{CD}$, and due to the custodial symmetry $SU(2)_{\rm cust}$ we have $f^{\pm}=f^0=f_{CD}$. The $W_\mu^{\pm}$ now have a mass:
\begin{equation}
M^2_W=\frac{1}{4}g^2f^2_{CD},
\end{equation}
where $f_{CD}\sim$ 246 GeV.

A mass matrix describes the $2\times 2$ matrix for the Abelian $B_\mu$ and the $W^3_\mu$ components of the $W^a_\mu$ boson:
\begin{equation}
M^2=\frac{1}{4}\begin{pmatrix}
g^{2} & gg'\\
gg' & g^{'2}
\end{pmatrix}f^2_{\rm C},
\end{equation}
where $g$ and $g'$ are the $SU(2)_L$ and $U(1)_Y$ coupling constants, respectively. The matrix eigenvalues are given by
\begin{equation}
M_\gamma^2=0,
$$ $$
M^2_Z=\frac{1}{4}(g^2+g^{'2})f_{CD}^2,
\end{equation}
where $M_\gamma$ and $M_Z$ denote the photon and $Z$ masses. We have 
\begin{equation}
\frac{M_W}{M_Z}=\frac{g}{(g^2+g^{'2})^{1/2}}=\cos\theta_w,
\end{equation}
where $\theta_w$ is the weak mixing angle. The standard tree level result for the $\rho$ parameter is
\begin{equation}
\rho=\frac{M_W^2}{M_Z^2\cos^2\theta_w}=1.
\end{equation}
The spectrum of composite quarks is determined and the quark masses are directly proportional to $\Lambda_{\rm CD}$. The quark spectrum is the same as in the SM, so we do not anticipate any conflict with the stringent bounds on flavor changing neutral currents and precision EW data.

As in QCD in which the masses of hadrons are generated dynamically, the masses of the quarks and leptons in our composite model will be generated dynamically. In contrast, in the SM the masses of the quarks and leptons are generated by their Yukawa couplings to the Higgs boson. In the SM the masses of fermions are parameterized by their associated coupling constants. {\it This fits the data but no explanation is given for the magnitudes of the particle masses.} 

\section{Conclusions}

The economical model proposed here suggests that the final basic building block of matter is the two-component doublet $\psi=\begin{pmatrix}\alpha\\\beta\end{pmatrix}$.
We can speculate {\it that the basic constituents of matter based only on spin 1/2 fermions can no longer be reduced to a further substructure}, for the next lower level of group structure is the Abelian $U(1)$.  The model has a scale invariant chiral symmetry for the massless constituents $\alpha$ and $\beta$ in analogy to the chiral symmetry of QCD. The scale invariance is broken by the confining energy scale $\Lambda_{CD}$. We can choose two ways to spontaneously break the $SU(2)_L\times U(1)_Y\rightarrow U(1)_{\rm EM}$, and give masses to the $W^{\pm}$ and $Z^0$ bosons keeping the photon massless.  The first way follows the usual SM Higgs boson spontaneous symmetry breaking procedure with $v_H\neq 0$. The second way promotes the condensate $v_{CD}\neq 0$ without the Higgs scalar field $\phi$ generating the $W^\pm$ and $Z^0$ masses. Instead, three massless Goldstone bosons form the longitudinal components of the massive $W^\pm$ and $Z^0$. 

The masses of the quarks are generated by the $SU(2)_{\rm C}$ confining dynamics with the triplet of hypergluons interacting with the $\alpha$ and $\beta$ constituent geminis. The masses of the quarks and the point-like leptons are either determined by the SM Yukawa interaction with the scalar Higgs boson $\phi$, or by the alternative picture that particle quantum field dynamics involving self-energy integral equations determine the masses of the constituent geminis and the leptons, and in turn the masses of the SM quarks~\cite{Moffat}. The gemini-hypergluon confining force satisfies an $SU(2)_C$ color charge $\alpha_{CD}$ screening and asymptotic safety, leading to a UV complete gauge theory and a resolution of the gauge hierarchy problem. The composite nature of the Higgs boson $H^0$ resolves the Higgs mass hierarchy problem. For energy values $E\lesssim E_c$, the constituent geminis are tightly confined, so that the SM quarks and $W^\pm$ and $Z$ are point-like. Moroever, the $\alpha$ and $\beta$ do not contribute to the loop calculations at lower energies. 
The standard theory of spontaneous symmetry breaking and the generation of fermion masses from the Yukawa fermion-Higgs Lagrangian with $\langle 0\vert\phi\vert 0\rangle\neq 0$ is treated as an effective infrared low energy model. The troublesome scalar Higgs mass and gauge hierarchy problems are removed by the composite model, allowing for a model of strong and weak interactions that satisfies the principle of naturalness.

\section{Acknowledgements}

This research was supported in part by Perimeter Institute for Theoretical Physics. Research at Perimeter Institute is supported by the Government of Canada through the Department of Innovation, Science and Economic Development Canada and by the Province of Ontario through the Ministry of Research, Innovation and Science.


\begin{thebibliography}{100}

\bibitem{Moffat} J. W. Moffat, J. Mod. Phys. {\bf A30}, 1550014 (2015), arXiv:1401.3029 [hep-ph].

\bibitem{Fritzsch} H. Fritzsch, arXiv:1604.05818 [hep-p].

\bibitem{Wilson} K. G. Wilson, Phys. Rev. {\bf D3}, 1818 (1971).

\bibitem{tHooft} G. t'Hooft, in Proceedings of the 1979 Cargese Institute on Recent Developments in 
Gauge Theories, p. 135, Plenum Press, New York 1980.

\bibitem{Branchina} V. Branchina and E. Messina, Phys. Rev. Lett. {\bf 111}, 241801 (2013) [arXiv:1307.5193 [hep-ph]].

\bibitem{Gross} D. J. Gross and F. Wilczek, Phys. Rev. Lett. {\bf 30}, 1343 (1973).

\bibitem{Politzer} H. D. Politzer, Phys. Rev. Lett. {\bf 30}, 1346 (1973).

\bibitem{Susskind} E. Farhi and L. Susskind, Physics Reports, {\bf 74}, 277 (1981).

\end{thebibliography}
\end{document}